# Geostatistical Rock Physics AVA Inversion

Leonardo Azevedo[1*], Dario Grana[2], Catarina Amaro[1]

[1] CERENA, DECivil, Instituto Superior Técnico, Lisbon University, Av. Rovisco Pais, 1049-001 Lisbon, Portugal. e-mail: leonardo.azevedo@tecnico.ulisboa.pt; catarina.amaro@tecnico.ulisboa.pt

[2] Department of Geology and Geophysics, University of Wyoming, 1000 E. University Ave., Laramie WY, 82071, U.S.A.. e-mail: dgrana@uwyo.edu

* Corresponding Author

Phone: (+351) 218417425

e-mail: leonardo.azevedo@tecnico.ulisboa.pt

# SUMMARY

Reservoir models are numerical representations of the subsurface petrophysical properties such as porosity, volume of minerals and fluid saturations. These are often derived from elastic models inferred from seismic inversion in a two-step approach: first, seismic reflection data are inverted for the elastic properties of interest (such as density, P-wave and S-wave velocities); these are then used as constraining properties to model the subsurface petrophysical variables. The sequential approach does not ensure a proper propagation of uncertainty throughout the entire geo-modelling workflow as it does not describe a direct link between the observed seismic data and the resulting petrophysical models. Rock physics models link the two domains. We propose to integrate seismic and rock physics modelling into an iterative geostatistical seismic inversion methodology. The proposed method allows the direct inference of the porosity, volume of shale and fluid saturations by simultaneously integrating well-logs, seismic reflection data and rock physics model predictions. Stochastic sequential simulation is used as the perturbation technique of the model parameter space, a calibrated facies-dependent rock physics model links the elastic and the petrophysical domains and a global optimizer based on cross-over genetic algorithms ensures the convergence of the methodology from iteration to iteration. The method is applied to a 3D volume extracted from a real reservoir dataset of a North Sea reservoir and compared to a geostatistical seismic AVA.

# 1      INTRODUCTION

The aim of seismic reservoir characterization studies is to infer the spatial distribution of the petro-elastic properties of the subsurface in the region of interest (Doyen 2007). This is frequently done using a sequential approach: first, seismic reflection data are inverted into elastic properties, such as impedances or velocities (Tarantola 2005); then, elastic properties are converted into petrophysical properties (e.g., porosity, volume of shale, saturations) through a rock physics (or petrophysical) inversion (Mukerji et al. 2001; Coléou et al. 2005; Doyen 2007; Bosch et al. 2010; Grana & Della Rossa 2010; Rimstad & Omre 2010; Kemper & Gunning 2014; Grana et al. 2017a).

Seismic inversion consists of characterizing the model parameter values (i.e., the subsurface elastic properties) of the physical system under investigation based on recorded seismic reflection data. Mathematically, the seismic inversion problem can be expressed as:

$$\mathbf{m} = \mathbf{F}^{-1}(\mathbf{d}_{\text{obs}}) + \mathbf{e}, \qquad (1)$$

where $\mathbf{F}$ is the forward operator through which the recorded seismic amplitudes $\mathbf{d}_{\text{obs}} \in \mathbf{R}^d$ are obtained from an elastic subsurface model $\mathbf{m} \in \mathbf{R}^m$, and $\mathbf{e}$ represents the error term associated with the observations and modelling uncertainties error present in the data. The solution can be expressed by a probability distribution, by a statistical estimator or by a set of multiple realisations of model parameters that fit the observed seismic data equally (Tarantola 2005). $\mathbf{F}$ is frequently approximated by the convolution of an input source pulse represented by a wavelet with the subsurface reflectivity coefficients computed from the subsurface elastic properties. In seismic inversion problems the intrinsic limitations of the geophysical method due to measurement errors and uncertainties in the model and observations (e.g., the confined bandwidth and resolution of the seismic data, noise and physical assumptions associated with the forward models) imply a non-linearity, non-uniqueness and ill-conditioning in the retrieved inverse solution (Tarantola 2005).

There are two main categories of approaches to solve the seismic inverse problem: deterministic or statistical (Bosch et al. 2010). Deterministic approaches are based on regression models of optimization algorithms aiming for a single best-fit solution. Such methods generally lack of a reliable assessment of the uncertainty associated with the retrieved subsurface model. Within this framework, the uncertainty may be assessed as a linearization around the best-fit inverse solution, which is normally retrieved by least squares, and in this sense, the uncertainty is strictly represented by a local multivariate Gaussian (Tarantola 2005).

Statistical approaches are able to retrieve the best-fit inverse model and allow the assessment of the uncertainty associated with the retrieved petro-elastic models. In statistical methods for seismic inversion , the solution is expressed as a probability density function in the model parameters space. This distribution allows quantifying the uncertainty related to the experimental data and the physical process being modelled (Tarantola 2005). Within the probabilistic setting, there are two main settings used to solve the seismic inverse problem: Bayesian linearized methods (e.g., Buland & Omre 2010; Grana & Della Rossa 2010; Grana et al. 2017a) and stochastic (e.g., González et al. 2008; Bosch et al. 2009; Azevedo & Soares 2017).

Stochastic or geostatistical seismic inversion methodologies are iterative procedures based on different stochastic optimization algorithms, e.g. simulated annealing, genetic algorithms, probability perturbation method, gradual deformation and neighborhood algorithm (e.g. Sen & Stoffa 1991; Sambridge 1999; Le Ravalec-Dupin & Nœtinger 2002; Soares et al. 2007; González et al. 2008; Grana et al. 2012; Azevedo et al. 2015; Azevedo & Soares 2017; Azevedo et al. 2018). With respect to Bayesian linearized approaches (e.g. Buland & Omre, 2003; Grana & Della Rossa 2010; Grana et al. 2017a), stochastic methods do not assume any parametric a priori distribution, and consequently are able to search the model parameter space more exhaustively. However, they are computationally expensive since analytical solutions are not available.

Stochastic inversion methods have been previously applied to seismic inversion problems for the estimation of elastic properties. However, such methods could be extended to facies classification and petrophysical property estimation. In rock physics modelling, petrophysical properties, such as porosity, volume of shale and saturations, are transformed into elastic attributes, using empirical relationships or more complex physical models based on poro-elasticity theory (Avseth et al. 2005; Mavko et al. 2009; Dvorkin et al. 2014).

In this work, we propose an iterative geostatistical seismic AVA inversion methodology to invert pre- or partially stacked seismic reflection data directly for porosity, volume of shale, water saturation, and facies by integrating rock physics modelling into the inversion procedure. At each iteration, the model parameter space, in the rock property domain, is perturbed by stochastic sequential simulation and co-simulation. The resulting petrophysical models are classified into facies using Bayesian classification. A calibrated facies-dependent rock physics model is applied in the forward model in order to derive the corresponding elastic properties. The rock physics model links the petrophysical and the elastic domains allowing the calculation of the synthetic seismic reflection data. Based on the trace-by-trace match between synthetic and real seismic data a new set of petrophysical models is generated in order to guarantee the converge of the methodology and match the real seismic dataset. In the Methodology section, we describe each step of the proposed methodology in detail. Then, in the Application section, we show the results of its application to a sector of a North Sea field. The results are compared to the results obtained by a geostatistical seismic AVA inversion (Azevedo et al. 2018).

## 2  METHODOLOGY

The proposed methodology is divided into four main steps: i) petrophysical model generation of porosity ($\phi$), volume of shale ($Vsh$) and water saturation ($Sw$) by stochastic sequential simulation and co-simulation; ii) facies classification based on the models generated in i); iii) elastic property calculation by applying a pre-calibrated facies-dependent rock physics model and synthetic seismic calculation; iv) comparison between real and synthetic seismic data and generation of new petrophysical models for the next iteration constrained by data mismatch computed in iii).

## 2.1 Petrophysical model generation

The first step consists in sequentially generating an ensemble of $Ns$ models of water saturation, porosity and volume of shale using stochastic sequential simulation and co-simulation. Direct Sequential Simulation (DSS; Soares 2001) and Direct Sequential co-Simulation (co-DSS) with joint probability distributions (Horta & Soares 2010) are used as the model perturbation technique. Differently from Sequential Gaussian Simulation (SGS; Deutsch & Journel, 1998), the use of DSS allows the direct use of the distribution of the property to be simulated as estimated form the experimental data, i.e. the well-log data. This is of interest when the distribution retrieved from the experimental well data is not Gaussian.

Within the iterative procedure, we use DSS to simulate (in iteration 1) and co-simulate (in the following iterations) water saturation models. Co-DSS with joint probability distributions is used to generate models of porosity and volume of shale. Porosity is co-simulated using the previously simulated water saturation as secondary variable and volume of shale is conditioned to the previously simulated porosity model. This stochastic sequential co-simulation technique ensures the reproduction of the relationship between a secondary variable and the variable to be simulated as inferred from the experimental data. To retrieve reliable petrophysical models we need to ensure that their relationship, as estimated from the well-log data, is reproduced in the inverse models. All the models generated during the inversion, reproduce the values of the well-log data at its location, the marginal and joint distribution of each property and a spatial continuity pattern expressed by a variogram model. The variogram models are fitted to experimental variograms computed from the existing well data.

## 2.2 Facies Classification

The models generated in the first step are classified into $Ns$ facies models. We use a Bayesian classification approach (Avseth et al. 2005) based on a training data built using well-log data and the collocated well-log facies profile. Alternative classification methods, such as Expectation-Maximization (Hastie et al. 2002), can be used. Grana et al. (2017b) compares the performance of both classification methods in seismic inversion.

At the well location, the well-log curves of porosity, volume of shale and water saturation are interpreted in well-log facies profile and classified into a given number of $N_f$ facies. Examples of facies include shale, brine sand and hydrocarbon sand. The statistical properties (i.e., mean, covariance and proportions) as inferred from the training data are used to compute the prior and likelihood function for the Bayesian classification according to Bayes' rule:

$$P(k|\boldsymbol{d}) = \frac{P(\boldsymbol{d}|k)P(k)}{P(\boldsymbol{d})} = \frac{P(\boldsymbol{d}|k)P(k)}{\sum_{k=1}^{N_f} P(\boldsymbol{d}|k)P(K)}, k = 1, \dots, N_f, \qquad (2)$$

where, $\boldsymbol{d}$ is the vector of the petrophysical properties used for the classification, the previously simulated petrophysical models, and $k$ is the facies value. In Equation 2, $P(\boldsymbol{d}|k)$ is the likelihood function, $P(k)$ is the prior model and $P(\boldsymbol{d})$ is a normalization constant.

## 2.3 Rock Physics modeling

Rock physics models link the reservoirs rock properties to their corresponding elastic response (Mavko et al. 2009). Different rock physics models differ in terms of assumptions and complexities of the physical formulation and derivation. Rock physics models include simple empirical relations inferred from laboratory experiments as well as more complex physical models based on Hertz-Mindlin contact theory or inclusion theory (Avseth et al. 2005; Mavko et al. 2009; Dvorkin et al. 2014).

Based on the available well-log data we first calibrate a rock physics model that fits the observed data within each facies. Indeed, many rock physics models include empirical parameters that should be chosen to fit the well-log measurements. Several rock physics models can be used; however, the general rock physics workflow aims at estimating the elastic response including density ($\rho$), P-wave velocity ($V_P$) and S-wave velocity ($V_S$) when the petrophysical properties are known. Elastic properties can be computed by definition as:

$$V_P = \sqrt{\frac{K_{sat} + \frac{4}{3}\mu_{sat}}{\rho}}; \qquad (3)$$

$$V_S = \sqrt{\frac{\mu_{sat}}{\rho}}; \qquad (4)$$

and

$$\rho = \rho_m(1-\phi) + \rho_{fl}\phi,$$

where $K_{sat}$ is the saturated-rock bulk modulus, $\mu_{sat}$ is the saturated-rock shear modulus, where $\rho_m$ is the matrix density, and $\rho_{fl}$ is the fluid density. The saturated-rock elastic moduli $K_{sat}$ and $\mu_{sat}$ are generally computed using Gassmann's equations. The saturated-rock bulk modulus depends on the dry-rock bulk modulus $K_{dry}$, the matrix bulk modulus $K_{mat}$, the fluid bulk modulus $K_{fl}$, and porosity $\phi$. The saturated-rock shear modulus is equal to the dry-rock shear modulus $\mu_{dry}$. The matrix, fluid, and dry properties can be computed according to different rock physics relations. In our formulation, we used the stiff sand model (Appendix A).

Once the elastic response is calculated, angle-dependent synthetic reflection seismic data can be obtained by computing angle-dependent reflection coefficients ($R_{pp}(\theta)$) using 3-term Shuey's linear approximation (Shuey 1985):

$$R_{pp}(\theta) \approx R(0) + G\sin^2\theta + F(\tan^2\theta - \sin^2\theta), \qquad (5)$$

where

$$R(0) \approx \frac{1}{2}\left(\frac{\Delta V_p}{\overline{V_p}} + \frac{\Delta \rho}{\bar{\rho}}\right);$$

$$G = R(0) - \frac{\Delta \rho}{\bar{\rho}}\left(\frac{1}{2} + \frac{2\Delta V_s^2}{\overline{V_p^2}}\right) - \frac{4\overline{V_s^2}}{\overline{V_p^2}}\frac{\Delta V_s}{\overline{V_s}};$$

$$F = \frac{1}{2}\frac{\Delta V_p}{\overline{V_p}},$$

where $\Delta V_P$, $\Delta V_S$ and $\Delta \rho$ are the differences of P-wave velocity, S-wave velocity and density above and below a reflection interface and $\overline{V_P}$, $\overline{V_S}$ and $\bar{\rho}$ are the corresponding average values across the interface. The resulting reflection coefficients are convolved with angle-dependent wavelets in order to generate synthetic seismic data.

## 2.4 Stochastic perturbation

The stochastic perturbation of the petrophysical models is based on a data selection procedure, performed at the end of each iteration and based on the principle of crossover genetic algorithm. The best portions of the petrophysical models generated at a given iteration, i.e. the regions of the model with the highest match between real and synthetic seismic data, are used as seed for the generation of a new family of models during the next iteration.

After generating the angle-dependent synthetic seismic, each synthetic and real traces with a total number of $N$ samples are compared and the mismatch is evaluated according to the following distance:

$$QC = \frac{2*\sum_{s=1}^{N}(x_s*y_s)}{\sum_{s=1}^{N}(x_s)*\sum_{s=1}^{N}(y_s)}, \qquad (6)$$

where $x$ and $y$ are the real and synthetic seismic traces respectively.

At each iteration, from the ensemble of rock properties models generated during the first stage of the proposed algorithm, we select the traces that ensure the highest match for all the angles simultaneously. These traces are used as secondary variables in the co-simulation of a new set of models during the following iteration.

The sampling and optimization algorithms in geostatistical rock physics seismic AVA (Figure 1) inversion can be summarized in the following sequence of steps:

i) Stochastic sequential algorithm (DSS; Soares 2001) for the generation of *Ns* water saturation models conditioned from the available well-log data and to a spatial continuity pattern as revealed by a variogram model;

ii) Stochastic sequential co-simulation with joint probability distributions (Horta & Soares 2010) for the generation of *Ns* models of porosity models given the *Ns* previously simulated water saturation models, the available porosity-log data and to a spatial continuity pattern as revealed by a variogram model;

iii) Stochastic sequential co-simulation with joint probability distributions (Horta & Soares 2010) for the generation of *Ns* models of volume of shale models given the *Ns* previously simulated porosity models and the available log of volume of shale and to a spatial continuity pattern as revealed by a variogram model;

iv) Litho-fluid facies classification of the rock property models generated in i), ii) and iii) into a number of pre-defined facies using Bayesian classification;

v) Application of facies-dependent rock-physics models to compute the elastic response of the models generated in i), ii) and iii) in each facies;

vi) Computation of angle-dependent reflection coefficients and convolution with angle-dependent wavelets to generate synthetic seismic data using Shueys' linear approximation;

vii) Comparison on a trace-by-trace basis of synthetic and real seismic data;

viii) Selection from the models generated in i), ii) and iii) of the portions that ensure the highest correlation coefficient simultaneously for all the angles at each reservoir location;

ix) Generation of a new set of rock property models by using co-simulation with local correlation coefficients.

x) Iteration until a given global correlation coefficient between the angle-dependent synthetic and real seismic data is reached.

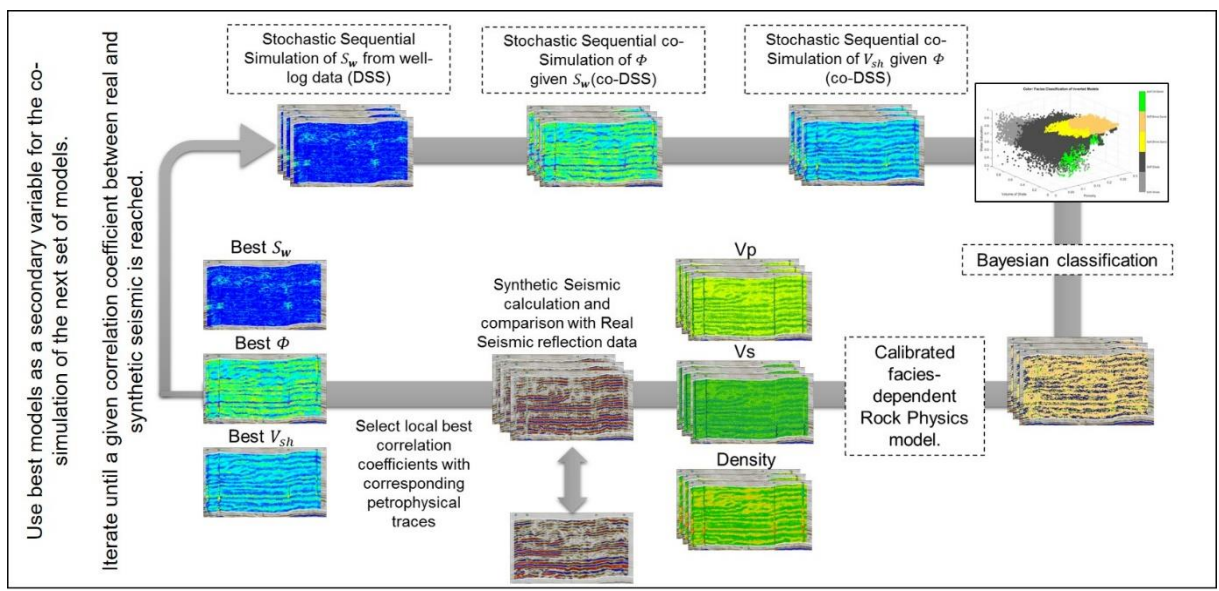

Figure 1 - Schematic representation of the proposed workflow for geostatistical rock physics seismic AVA inversion.

## 3 REAL-CASE APPLICATION

We applied the geostatistical inversion to a subvolume of seismic data extracted from the segment C of the Norne field dataset. The available dataset includes three partial angle stacks acquired in 2001 with central angles of incidence of 10º, 22.5º and 35º (i.e., near-stack, mid-stack and far-stack, respectively) and four wells (Figure 2). The inversion grid is defined by 109 by 79 cells in inline and crossline directions and 75 samples in the vertical direction. The sampling rate is 4 ms. Angle-dependent wavelets were statistically extracted for each partial stack using standard commercial software.

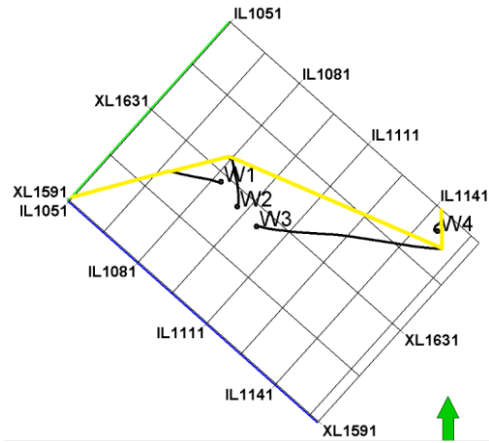

Figure 2 - Seismic grid and location of the four wells used as constraining data for the inversion. Yellow line shows the location of the vertical well section used to show the results of the inversion.

Two out of the four wells (wells W1 and W3) have a complete log suite composed by P-wave and S-wave velocity, density, water saturation, volume of shale and porosity logs (Figure 3). Well-logs were used for well-log facies classification, rock physics calibration and as conditioning data for the generation of the reservoir models of water saturation, porosity and volume of shale. Three facies have been defined: shale with volume of shale higher than or equal to 40 %, brine sand with volume of shale smaller than 40% and water saturation higher than or equal to 80% and oil sand with volume of shale smaller than 40% and water saturation smaller than 80% (Suman and Mukerji 2013). As in Grana and Mukerji (2015), the elastic response in sandstone was modeled with the stiff-sand model. A mixture of quartz and feldspar ($K_{sa} = 25\ GPa$, $G_{sa} = 20\ GPa$, $\rho_{sa} = 2.64\ g/cm^3$) was considered for the sandstone phase. The matrix bulk and shear moduli were computed using Voigt-Reuss-Hill average. The critical porosity set as 0.49 and the effective fluid bulk modulus was computed using Batzle-Wang relation (Batzle and Wang 1992) and Reuss average. Gassmann's equation was applied to compute the fluid effect and predict the elastic behaviour in wet conditions. Han's model empirically calibrated using well-log data at the well location (e.g. Mavko et al. 2009) was used to model the elastic response in shale. The existing well-log samples of water saturation, porosity and volume of shale for wells W2 and W4 were used exclusively as conditioning data for the stochastic sequential simulation and co-simulation. The original well-logs of water saturation, porosity and volume of shale were upscaled into the inversion grid and used to compute the experimental variograms to model the spatial continuity pattern.

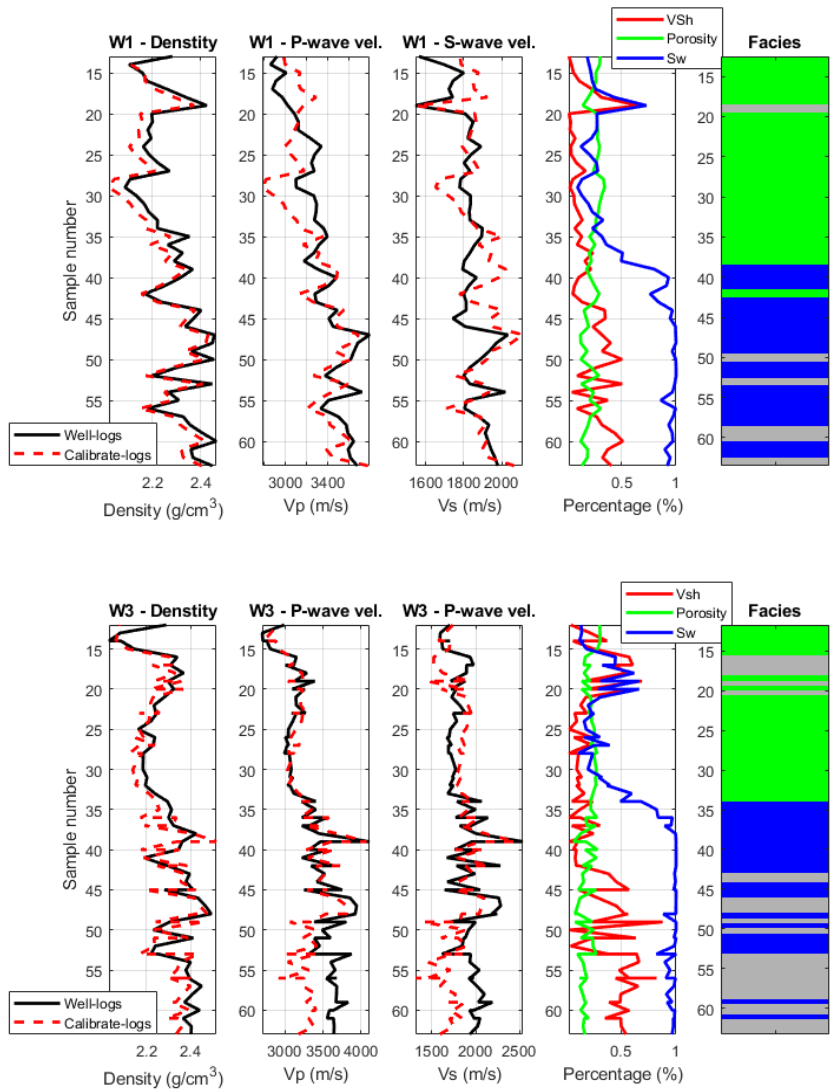

Figure 3 – Comparison between well-log samples of density, $V_P$ and $V_S$ and samples predicted using the facies-dependent calibrated rock physics models for wells W1 and W3. Upscaled petrophysical-log samples used as input for the rock physics model are shown in the fourth track. Fifth track shows the manual facies classification: samples classified as shales are plotted in gray, as brine sands in blue and as oil sands in green.

We ran the inversion with six iterations with the generation of thirty-two realisations of water saturation, porosity and volume of shale per iteration. The iterative procedure converges until it reaches 66% of correlation coefficient between the synthetic and real seismic volumes for all angles simultaneously (Figure 4). As the perturbation of the model parameter space is done with stochastic sequential simulation and co-simulation, the histograms of water saturation, porosity and volume of shale as inferred from the upscaled-log samples are reproduced in all models generated during the inversion (Figure 5). The reproduction of the joint distribution of water saturation and porosity and porosity and volume of shale is also ensured by using direct sequential co-simulation with joint probability distributions (Figure 6) (Horta & Soares 2010).

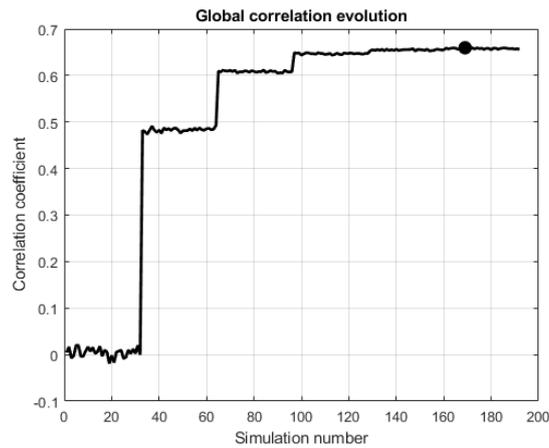

Figure 4 – Evolution of the global correlation coefficient between synthetic and real seismic simultaneously for all partial stacks. Black filled circle represents the best-fit inverse models (iteration 6, simulation 9).

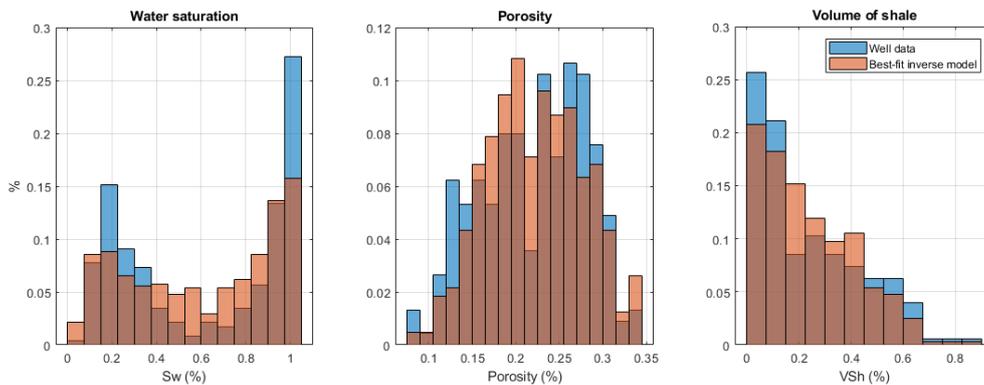

Figure 5 – Comparison between histograms from upscaled well-log samples and best-fit inverse petrophysical models.

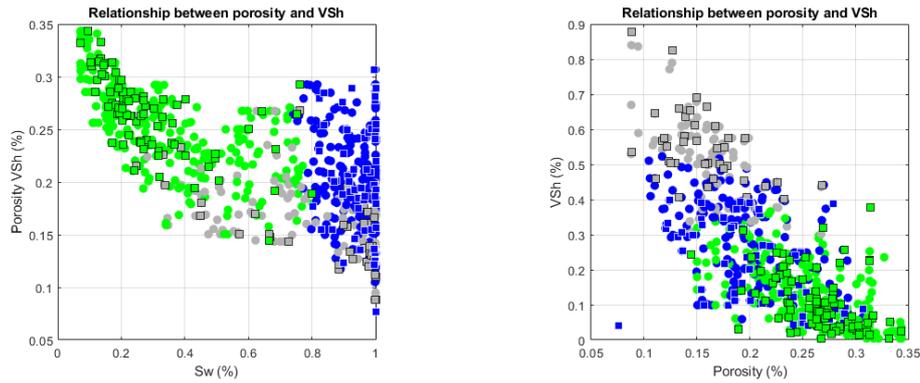

Figure 6 – Comparison between joint distributions of water saturation vs. porosity and porosity vs. volume of shale. Upscaled-log samples are plotted in squares while circles represent the best-fit inverse petrophysical models. Samples classified as shales are plotted in gray, as brine sands in blue and as oil sands in green.

The well-log facies classification of the upscaled-logs is also reproduced. Observable misclassifications can be due to the probabilistic nature of the classification method used in the proposed methodology and the stochasticity of the petrophysical models generated at each iteration.

The real partial angle stacks show a bright reflection around the 2400 ms corresponding to the top of the Not formation (Figure 7). The spatial continuity of the reflections decreases with depth, which makes the convergence of the inversion in these areas more challenging. Below 2600 ms it is hard to identify any spatially coherent reflection. As the incidence angle increases, the number of continuous reflections decreases. The far-stack volume (Figure 7c), shows a single clear reflection corresponding to the top of the Not formation.

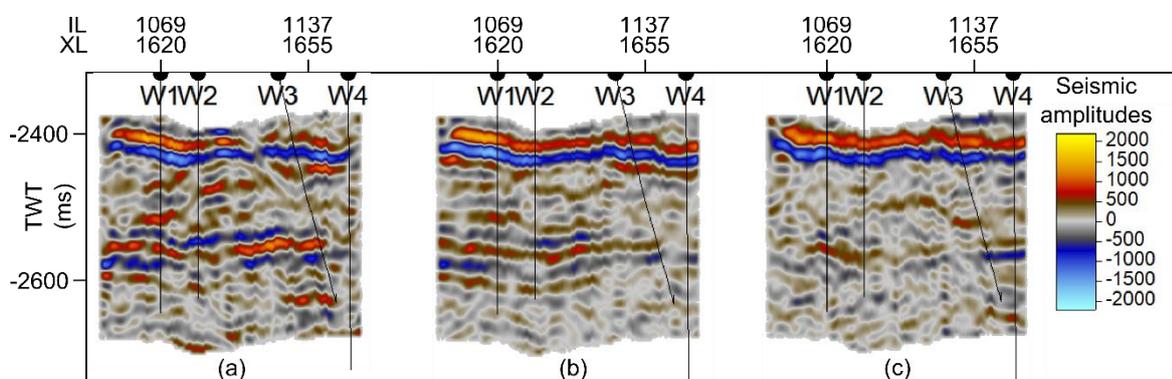

Figure 7 – Vertical well sections extracted from the real partial angle stacks: (a) near-stack; (b) mid-stack; and (c) far-stack.

The synthetic seismic reflection data computed from the elastic models predicted using the best-fit inverted water saturation, porosity and volume of shale models is shown in Figure 8. The synthetic predicts the location and amplitude of the main seismic reflections. However, it introduces higher reflections continuity compared to the observed seismic. This might be due the horizontal variogram ranges imposed during the simulation of the petrophysical models.

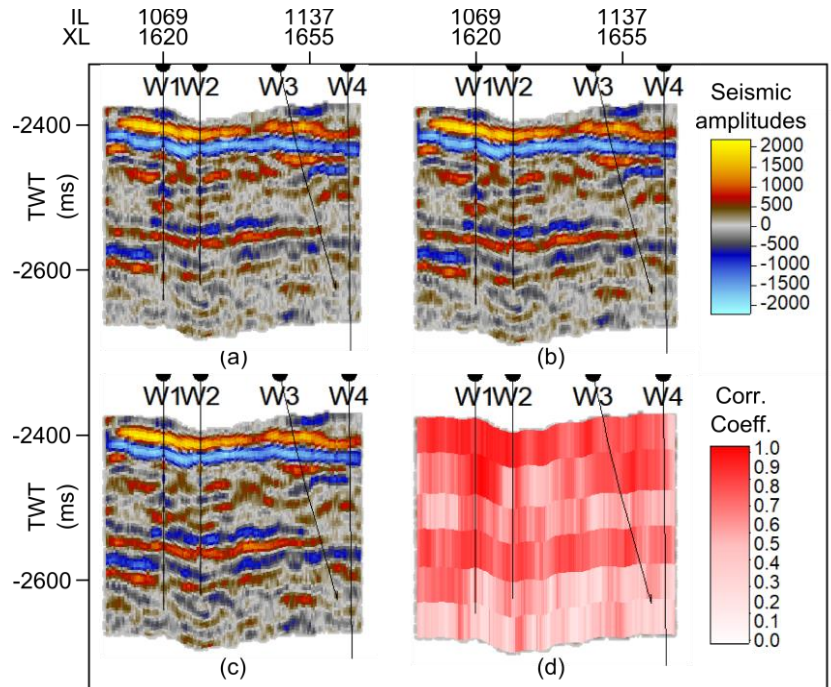

Figure 8 – Vertical well sections extracted from the best-fit synthetic partial angle stacks: (a) near-stack; (b) mid-stack; and (c) far-stack. (d) Shows the average local correlation coefficient between synthetic and real seismic for all three angles.

The difficulty in reproducing the bottom part of the seismic data (below 2600 ms) is shown by the lower trace-by-trace correlation coefficients between real and synthetic seismic traces. This metric clearly shows a better convergence in the top region of the seismic data and decreasing values in depth.

For illustration purposes, we show the average models of water saturation, porosity and volume of shale computed from all the realisations generated during the last iteration (Figure 9). Note that all models generated during the last iteration produce synthetic seismic with similar correlation coefficient (Figure 4). These models show a larger spatial variability in the upper part of the inversion model associated with the regions of lower water saturation and preserve with the geological features interpreted from the original seismic. The elastic response predicted by these models is shown in the bottom row of Figure 9.

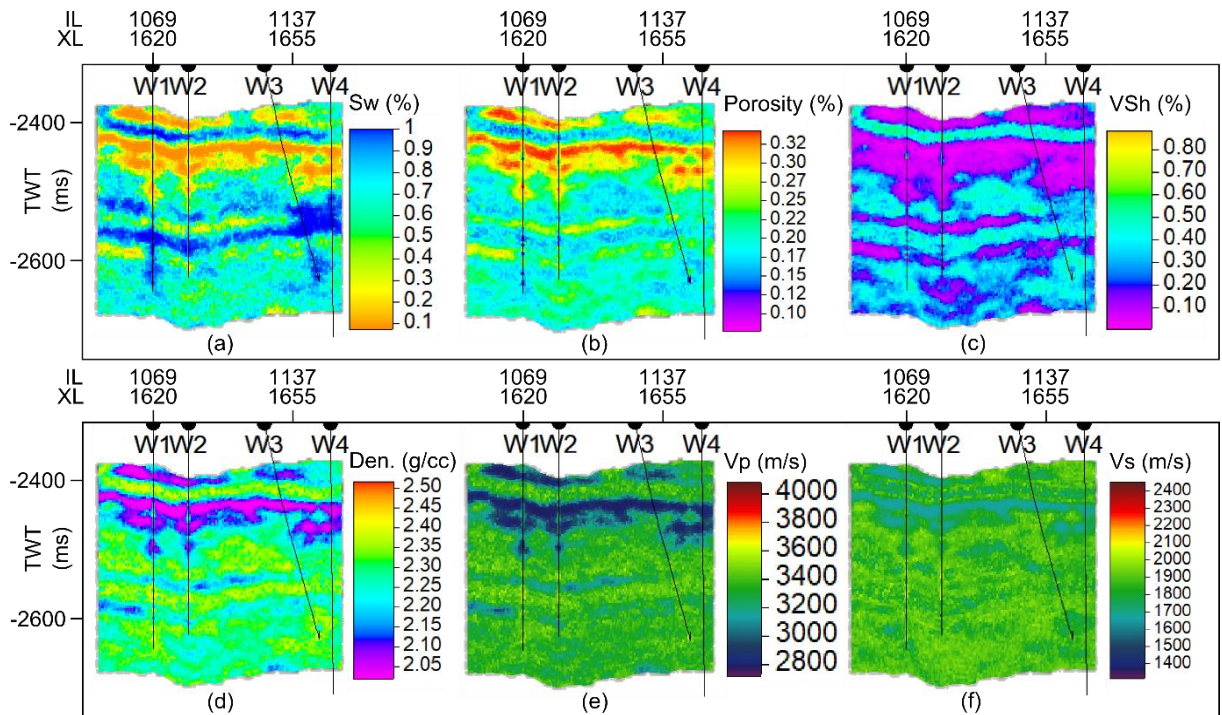

Figure 9 - Vertical well sections extracted from the average model of the realisations generated during the last iteration for: (a) volume of shale; (b) porosity; and (c) volume of shale. The corresponding predicted elastic models for: (d) density; (e) $V_P$; and (f) $V_s$ are shown below.

We use wells W4 and W2 to verify the reliability of the predicted elastic samples in the two wells not used for the calibration of the rock physics (Figure 10). We compare the existing measured log samples of density, P-wave and S-wave velocity to the minimum, average and maximum values generated during the last iteration of the geostatistical inversion. For well W4 (top row Figure 10), the density and $V_P$ logs agree with the average predict values and is mainly within the bounds of the minimum and maximum predicted values. Well W2 (bottom row Figure 10) is correctly constrained for almost the entire extent of the well where all the models generate the same water saturation, porosity and volume of shale and consequently the same elastic response. The most likely facies model computed from all the facies volumes generated during the last iteration do match the log facies interpretation (fifth plot in Figure 10).

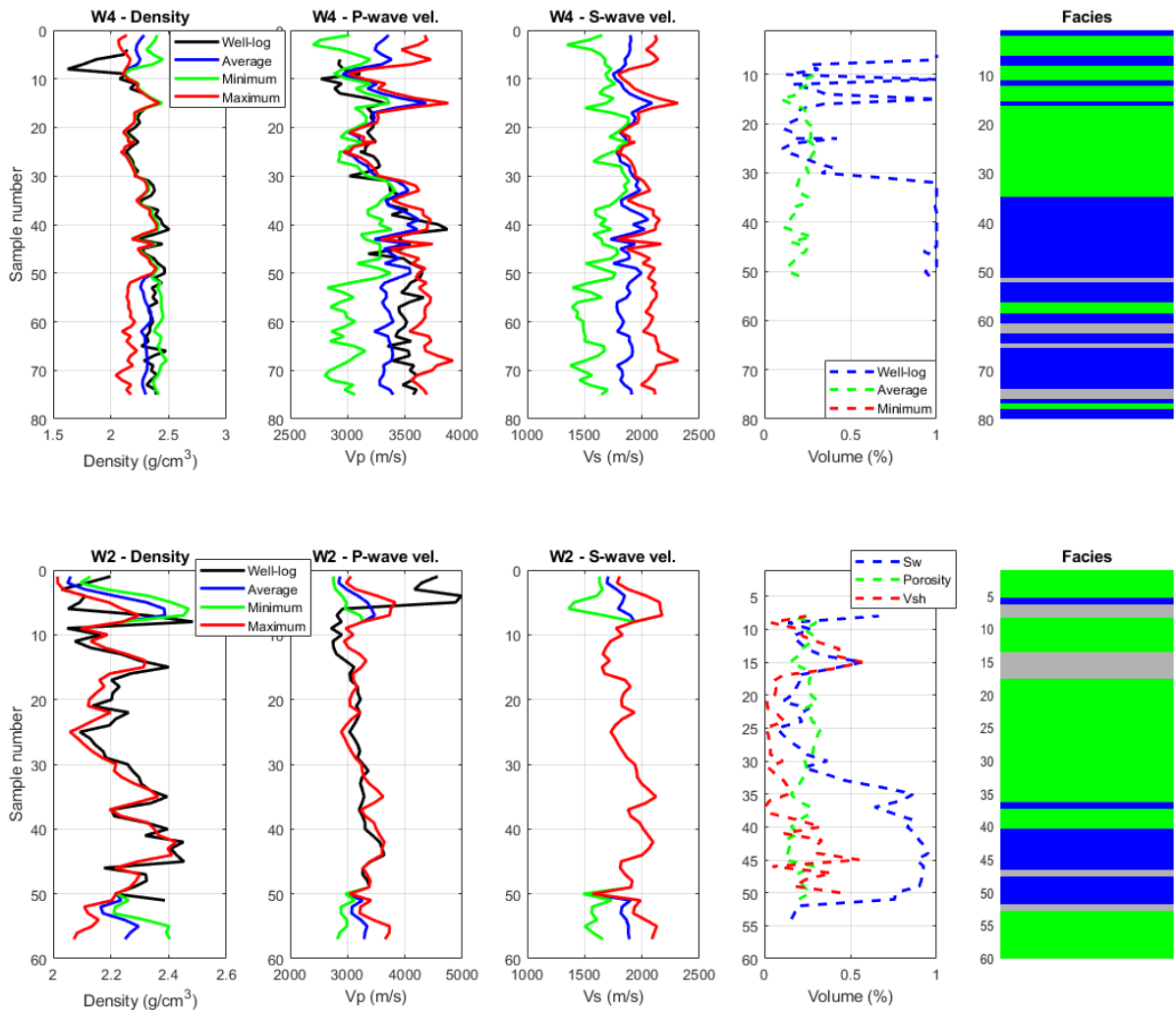

Figure 10 – Comparison between well-log and predicted minimum, average and maximum of density, $V_P$ and $V_S$ samples in the last iteration of the inversion. The upscaled-log samples used as conditioning data for both locations are shown in the fourth track, these do not cover the entire extent of the wells. The resulting facies classification is shown in the fifth track: samples classified as shales are plotted in gray, as brine sands in blue and as oil sands in green.

Furthermore, we show the facies probability volumes of shale, brine and oil sand (Figure 11). These are computed by using the ensemble of facies models classified from the water saturation, porosity and volume of shale models generated during the last iteration. It is possible to interpret a main oil sand region immediately below the top Not formation (~2400 ms) and a thinner one below the top Aare formation (~2600 ms).

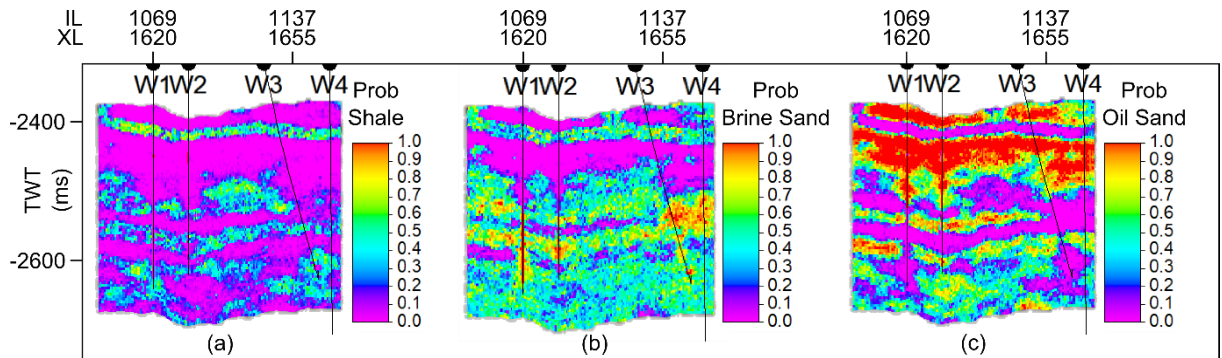

Figure 11 - Vertical well sections extracted from the probability volumes of: (a) shale; (b) brine sand; and (c) oil sand. These probabilities are computed from the ensemble of facies model generated during the last iteration.

Finally, we show the comparison between elastic models retrieved by a similar geostatistical seismic AVA inversion (Azevedo et al. 2018) where no rock physics modelling is used. The inversion was parameterized similarly to the application example shown for the proposed inversion. The same amount of iterations and realisations per iteration, reaching similar convergence rate at the end of the inversion. The average density, P-wave and S-wave velocity models of the last iteration are shown in Figure 12. While the location of the main features of interest is similar in the elastic models retrieved by both methods, there is a higher spatial variability on the models resulting from the geostatistical seismic AVA inversion. This is a direct consequence of using stochastic sequential simulation to generate these properties, contrary to derived them using a calibrated rock physics model.

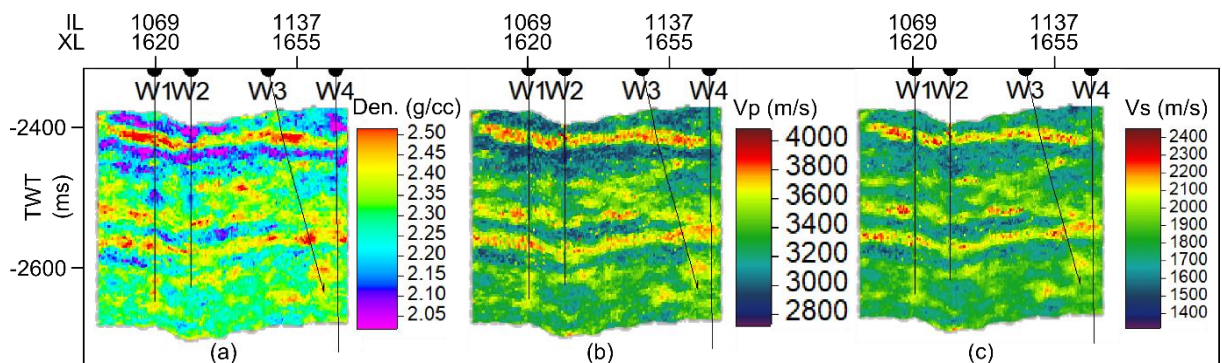

Figure 12 - Vertical well sections extracted from the average models of all realisations generated during the last iteration of the geostatistical seismic AVA inversion. (a) Density. (b) $V_P$. (c) $V_S$.

# 4    DISCUSSION

Our method is presented using stiff-sand model and Han's equation to model the elastic responses from water saturation, porosity and volume of shale for sand and facies respectively. However, it can be extended to any other model as neither the global perturbation technique nor the global optimizer depend on the specific rock physics model. The same applies for the classification technique used in the geostatistical petrophysical inversion. Here we use Bayesian classification as it is easy to implement and allows the use of training data built from a log-facies interpretation.

An important feature of the proposed method is the ability to reproduce the joint distribution between variables which improves the facies classification. This is ensured by using direct sequential co-simulation with joint probability distributions (Horta & Soares 2010). On the other hand, one of the potential limitations of the method is the need to define the order of the simulations of the property. Here, we decided to use a geostatistical criterion starting by the simulation of the property expected to be smother at the large-scale, water saturation.

As we are using a geostatistical inversion method, the modelling of experimental variograms might be an additional challenge. Often the number of wells within the region of interest is limited and wells are often located far from each other. This does not allow computing robust experimental variograms which results in inadequate variogram models that might impose biased continuities.

# 5    CONCLUSIONS

We propose a single-loop iterative geostatistical petrophysical seismic AVA inversion. First, petrophysical properties are simulated and co-simulated from the existing well-log data and classified in facies. Facies-dependent rock physics models are then applied to predict the elastic response. These elastic volumes are used to computed synthetic seismic data. The mismatch between synthetic and real seismic data drives the iterative inversion procedure. The final result is a set of updated models conditioned by seismic and well-log data. The proposed method is applied to a section of segment C of Norne dataset. The results show reliable petro-elastic models. Two wells not used in the calibration of the rock physics models show elastic logs within the predicted inverted bounds for the same properties.

# ACKNOWLEDGEMENTS

The authors would like to thank Statoil (operator of the Norne field) and its license partners ENI and Petoro for the release of the Norne data and the Center for Integrated Operations at NTNU for cooperation and coordination of the Norne Cases. The authors also acknowledge Schlumberger for the donation of the academic licenses of Petrel® and CGG for Hampson-Russell suite. LA and CA gratefully acknowledge the support of the CERENA (strategic project FCT-UID/ECI/04028/2013). CA acknowledges the Department of Geology and Geophysics of University of Wyoming for the six months internship.

*Author contributions*: LA and DG conceptualized and designed the method, and wrote the manuscript. LA and CA did the implementation of the method and ran the application example.

# APPENDIX A: ROCK-PHYSICS MODEL

The stiff-sand model based on Hertz-Mindlin grain-contact theory estimates bulk $K_{HM}$ (Equation A1) and shear $\mu_{HM}$ (Equation A2) moduli of a dry rock, under the assumption that the sand frame with a random pack of identical spherical grains is under an effective pressure $P$, with a critical porosity $\phi_c$ and a coordination number $n$ (i.e. average number of contacts per grain) (Mavko et al. 2009):

$$K_{HM} = \sqrt[3]{\frac{n^2(1-\phi_c)^2\mu_m^2 P}{18\pi^2(1-\nu)^2}};\qquad(A1)$$

and

$$\mu_{HM} = \frac{5-4\nu}{10-5\nu}\sqrt[3]{\frac{3n^2(1-\Phi_c)^2\mu_m^2 P}{2\pi^2(1-\nu)^2}}\qquad(A2)$$

where $\nu$ is the grain Poisson's ratio and $\mu$ is the shear modulus of the solid phase. In our implementation of the stiff-sand model, the matrix elastic moduli is calculated according to Voigt-Reuss-Hill averages for a matrix with sand and shale:

$$K_m = \frac{1}{2}\left(V_{sh}K_{sh} + (1-V_{sh})K_{sa} + \frac{1}{\frac{V_{sh}}{K_{sh}} + \frac{K_{sa}}{(1-V_{sh})}}\right)\qquad(A3)$$

and

$$\mu_m = \frac{1}{2}\left(V_{sh}\mu_{sh} + (1-V_{sh})\mu_{sa} + \frac{1}{\frac{V_{sh}}{\mu_{sh}} + \frac{\mu_{sa}}{(1-V_{sh})}}\right)\qquad(A4)$$

where $V_{sh}$ is the volume of shale, $K_{sh}$ and $\mu_{sh}$, are the bulk and shear moduli of shale, and $K_{sa}$ and $\mu_{sa}$, are the bulk and shear moduli of sand.

Assuming an isotropic linear elastic composite, and effective porosity values between zero and critical porosity $\phi_c$, this model connects the matrix elastic model ($K_m$ and $\mu_m$) of the solid-phase, with the elastic model of the dry rock ($K_{HM}$ and $\mu_{HM}$) at porosity $\phi_c$ using the with the modified Hashin-Shtrikman upper bound (Mavko et al. 2009):

$$K_{dry} = \left[\frac{\phi/\phi_c}{K_{HM} + \frac{4}{3}\mu_m} + \frac{1 - \phi/\phi_c}{K_m + \frac{4}{3}\mu_m}\right]^{-1} - \frac{4}{3}\mu_m, \qquad (A5)$$

and

$$\mu_{dry} = \left[\frac{\phi/\phi_c}{\mu_{HM} + \frac{4}{3}\xi\mu_m} + \frac{1 - \phi/\phi_c}{\mu_m + \frac{4}{3}\xi\mu_m}\right]^{-1} - \frac{1}{6}\xi\mu_m, \qquad (A6)$$

where

$$\xi = \frac{9K_m + 8\mu_m}{K_m + 2\mu_m}. \qquad (A7)$$

In presence of a fluid mixture or suspension, density is computed by using Woods' formula (Wood 1955). It assumes that the composite rock and each of its components are isotropic, linear and elastic.

$$\rho_m = \rho_{sh}V_{sh} + \rho_{sa}(1 - V_{sh}), \qquad (A8)$$

and

$$\rho_{fl} = \rho_w S_w + \rho_{oil}(1 - S_w), \qquad (A9)$$

where $\rho_{sh}, \rho_{sa}, \rho_w,$ and $\rho_{oil}$ are the densities of shale, sand, water and oil respectively and are assumed to be 2.59, 2.64, 1 and 0.81 g/cm³. The saturated bulk and shear moduli are are estimated using Gassmann's equations (Gassmann, 1951):

$$K_{sat} = K_{dry} + \frac{\left(1 - \frac{K_{dry}}{K_{mat}}\right)^2}{\frac{\phi}{K_{fl}} + \frac{1 - \phi}{K_m} - \frac{K_{dry}}{K_m^2}}, \qquad (A10)$$

where $\phi$ represents porosity, $K_{dry}$ and $K_{sat}$ denote the effective dry and saturated moduli of the rock, $K_m$ and $K_{fl}$ represent the matrix and fluid moduli respectively. The shear modulus is not sensitive to pore fluids:

$$\mu_{sat} = \mu_{dry}. \quad (A11)$$